\documentclass[aps,pre,onecolumn,showpacs,showkeys,amsmath,amssymb,
groupedaddress]{revtex4}

\usepackage{graphicx}
\usepackage{times}
\newcommand{\prt}{\partial}
\newcommand{\prm}{\prime}

\begin{document}

\preprint{}

\title{Standing and travelling waves in the shallow-water circular 
hydraulic jump}

\author{Arnab K. Ray}
\email{akr@iucaa.ernet.in}
\affiliation{Inter--University Centre for Astronomy and Astrophysics \\ 
Post Bag 4, Ganeshkhind, Pune University Campus \\ Pune 411007, India} 
\altaffiliation[Also at: ]{Harish--Chandra Research Institute, 
Chhatnag Road, Jhunsi, Allahabad 211019, India}

\author{Jayanta K. Bhattacharjee}
\email{tpjkb@mahendra.iacs.res.in}
\affiliation{Department of Theoretical Physics \\
Indian Association for the Cultivation of Science \\
Jadavpur, Kolkata 700032, India}

\date{\today}

\begin{abstract}
A wave equation for a time-dependent perturbation about the steady 
shallow-water solution emulates the metric an acoustic white hole, 
even upon the incorporation of nonlinearity in the lowest order. 
A standing wave in the sub-critical region of the flow is stabilised 
by viscosity, and the resulting time scale for the amplitude decay 
helps in providing a scaling argument for the formation of the hydraulic 
jump. A standing wave in the super-critical region, on the other hand,
displays an unstable character, which, although somewhat mitigated by
viscosity, needs nonlinear effects to be saturated. A travelling 
wave moving upstream from the sub-critical region, destabilises the 
flow in the vicinity of the jump, for which experimental support has 
been given. 
\end{abstract}


\maketitle

\section{Introduction}
\label{sec1}

The hydraulic jump is a subject of textbooks in fluid 
dynamics~\cite{wj,landau,granger,faber,ghpm}. With a jet of water impinging 
vertically on a flat surface, it can be seen that from the point of 
impingement, the water spreads out radially in a thin film, and beyond 
a certain distance, the height of the fluid layer abruptly increases, 
exhibiting the circular hydraulic jump~\cite{tan49,watson64,ot66}. This 
is a common 
occurrence in a kitchen sink, and somewhat differently this phenomenon 
is also exhibited as a tidal bore moving upstream in a river~\cite{faber}, 
which makes the hydraulic jump an engineering problem of considerable 
practical interest as well. 

Viscosity of the flowing liquid is the key factor that makes the hydraulic 
jump possible~\cite{bdp93,behh96,bpw97}, and in consequence of this, 
a scaling relation for the hydraulic jump radius has been  
derived by Bohr et al.~\cite{bdp93}, with strong 
experimental support coming subsequently in favour of this scaling 
relation through a work by Hansen et al.~\cite{hansen97}. Some later 
works have also shown that surface tension has an involvement with the 
jump in shallow-layer flows~\cite{bush03,rama06,rol07}. While dwelling
on these matters, 
it would be very pertinent to mention that in a recently 
published work, Rolley et al.~\cite{rol07} have brought forth an 
interesting experimental study in which the hydraulic jump was seen
to form in the flow of liquid helium, below its superfluid transition
temperature. The radius of the jump and the depth profile of the flow
were also not altered appreciably at the superfluid transition, as 
though the fluid was still normal. This feature is probably due to 
the fact that the flow velocity in this situation is supercritical
with respect to the critical velocity for superfluidity. 

While it is usual to address all jump-related issues in the 
stationary regime, in a departure from this practice, 
explicit time-dependence in the shallow-water approach to the circular 
hydraulic jump has been considered in this paper. A time-dependent 
perturbation has been imposed on the steady volumetric flow rate.
As an immediate and interesting consequence of 
this line of attack, a linearised equation of motion for the perturbation 
has been found to have a mathematical form that is very much like the 
metric of an acoustic white hole. This compatibility makes it possible 
to match the critical properties of the flow with the horizon condition 
of the acoustic white hole, with important consequences leading to
the formation of the jump itself. 

As regards the stability of stationary 
solutions, it has been shown that when the perturbation 
is treated as a standing wave, bounded at two separate spatial points
in the sub-critical flow region, viscosity will cause a damping 
of the amplitude of the perturbation.  
A resultant time scale for viscous damping of the perturbation has 
been argued to be the same time scale on which the flow is dissipatively
slowed down by viscosity, and using this time scale, along with the 
analogy of a white hole, it becomes possible to establish a 
physical explanation for the standard scaling relation of the hydraulic 
jump radius~\cite{bdp93}. A study to this effect has been reported 
before~\cite{sbr05} for the one-dimensional channel flow. 

In the super-critical region, however, the standing wave does not 
exhibit similarly stable behaviour. Viscosity goes some way in 
opposing the growth of the perturbation in this region, but this 
divergent aspect of the perturbation has to be saturated ultimately
by the conspicuously nonlinear effects in the super-critical region 
of the flow. This feature is in keeping with what has been seen from
earlier experiments~\cite{vol06}. A very intriguing feature that has
emerged on including nonlinearity is that there is no alteration of
the conditions under which it is possible to construct an acoustic 
metric for the flow. 
 
As opposed to standing waves, treating the perturbation as a travelling 
wave leads to a different conclusion about the stability of the
flow. In this case it can be seen that 
the outgoing mode of the travelling wave has a growing amplitude in 
the super-critical region of the flow, but it ultimately
decays out in the sub-critical flow region. On the other hand, the 
incoming mode causes great disturbance in the flow as it propagates 
upstream in the sub-critical region, especially very close to the 
jump. Experimental evidence has been adduced for the latter kind of
behaviour~\cite{kdc06}. 

\section{A wave equation and the metric of an acoustic white hole}
\label{sec2}

The flow is described by the Navier-Stokes equation and 
the equation of continuity, with both of them modified for 
a shallow radial flow confined to a plane~\cite{bdp93}. The steady 
flow variables would be the velocity of the flow, $v(r)$, and 
the local height of the incompressible fluid, $h(r)$, with the 
former having been
obtained in the shallow-water theory by vertically averaging the 
radial component of the velocity, $u(r,z)$, over the height, $h(r)$.
The boundary conditions for the averaging are $u(r)=0$ at 
$z=0$, and $\prt u/\prt z = 0$ at $z=h(r)$, under the assumption that 
while the vertical velocity is much small compared with the radial 
velocity, the vertical variation of the velocity (through the shallow 
layer of water) is much greater than its radial variation~\cite{bdp93}. 

The dynamic generalisation of both $v$ and $h$, and 
their corresponding dynamic equations may be written as 
\begin{equation}
\label{con}
\frac{\prt h}{\prt t} + \frac{1}{r}\frac{\prt}{\prt r}
\left( rvh \right ) = 0
\end{equation}
and 
\begin{equation}
\label{ns}
\frac{\prt v}{\prt t} + v\frac{\prt v}{\prt r} + g\frac{\prt h}
{\prt r} = - \nu \frac{v}{h^2} , 
\end{equation}
respectively, with the usual viscous term in the Navier-Stokes 
equation having been suitably approximated in Eq.~(\ref{ns}), following 
the arguments offered by Bohr et al.~\cite{bdp93}. 
The steady solutions of Eqs.~(\ref{con}) and~(\ref{ns}) above lead to 
the phenomenon of the hydraulic jump. 

To carry out a linearised perturbative analysis, it will be 
convenient to work with a new variable that is defined as
$f=rvh$. The wisdom behind this choice may be discerned from the 
structure of Eq.~(\ref{con}), the steady solution of which can be 
integrated to give $rvh = Q/2 \pi$, with the integration
constant $Q$ being the steady volumetric flow rate. Therefore, 
the new variable $f$ can be physically identified with the 
time-dependent volumetric flow rate, and its steady solution
will be a constant. 

Solutions of the form $v(r,t) = v_0(r)
+ v^{\prm}(r,t)$ and $h(r,t) = h_0(r) + h^{\prm}(r,t)$ are to
be used, in which the subscript $0$ indicates steady
solutions, while the primed quantities are time-dependent 
perturbations about those steady solutions. Linearising in these
fluctuating quantities gives the fluctuation of $f$ about its
constant steady value, $f_0 = r v_0 h_0 = Q/2 \pi$, as 
\begin{equation}
\label{flucf}
f^{\prm}=r \left( v_0 h^{\prm} + h_0 v^{\prm} \right) . 
\end{equation}
In terms of $f^{\prm}$, it is possible to write from Eq.~(\ref{con}),
\begin{equation}
\label{fluch}
\frac{\prt h^{\prm}}{\prt t} = - \frac{1}{r}
\frac{\prt f^{\prm}}{\prt r} , 
\end{equation}
which, in conjunction with Eq.~(\ref{flucf}), gives,
\begin{equation}
\label{flucv}
\frac{\prt v^{\prm}}{\prt t} = \frac{1}{rh_0}\left(\frac{\prt f^{\prm}}
{\prt t}\right) + \frac{v_0}{rh_0} 
\left(\frac{\prt f^{\prm}}{\prt r} \right) . 
\end{equation}
A further partial differentiation of Eq.~(\ref{flucv}) with respect 
to $t$, followed by extracting the linearised terms in $v^{\prm}$ and
$h^{\prm}$ from Eq.~(\ref{ns}), and substitution of these perturbed
quantities in terms of $f^{\prm}$, will deliver 
\begin{equation}
\frac{\prt}{\prt t} \left[v_0
\left( \frac{\prt f^{\prm}}{\prt t}\right)\right]
+ \frac{\prt}{\prt t} \left[v_0^2
\left( \frac{\prt f^{\prm}}{\prt r}\right)\right]
+ \frac{\prt}{\prt r} \left[v_0^2
\left( \frac{\prt f^{\prm}}{\prt t}\right)\right]
+ \frac{\prt}{\prt r} \left[v_0
\left(v_0^2 - gh_0 \right) \frac{\prt f^{\prm}}{\prt r}\right] 
= - \nu \frac{v_0}{h_0^2}\left(\frac{\prt
f^{\prm}}{\prt t} + 3 v_0 \frac{\prt f^{\prm}}{\prt r} 
\right) . 
\label{metric}
\end{equation}

It should be instructive to examine Eq.~(\ref{metric}) in its inviscid 
limit, i.e. when $\nu=0$.  
Some studies~\cite{su02,blv05,vol05} have revealed a 
close analogy between the propagation of a wave in a moving fluid and of 
light in curved space-time. On this particular issue, Sch\"{u}tzhold 
and Unruh have shown that gravity waves in a shallow layer of liquid 
are governed by the same wave equation as for a scalar field in 
curved space-time~\cite{su02}. For an inviscid, incompressible and 
irrotational flow, the flow velocity has been defined
to be the gradient of a scalar potential. Perturbing this potential 
about its background value, under the restricted condition of the 
background flow height being a constant, leads to
an effective metric, in which the velocity of gravity
waves replace the speed of sound in sonic analogues which closely
reflect features seen in general relativistic studies~\cite{su02,blv05}.

The present analysis is devoted to what is essentially a 
dissipative system 
(since it includes viscosity). However, in its inviscid limit, this 
system will deliver the same metric obtained by Sch\"{u}tzhold and Unruh 
from their purely inviscid model. It must be stressed here that the 
choice of perturbing the constant background flow rate
is paticularly expedient, since conservation of matter holds good
even in a system that undergoes viscous dissipation. It is to be
further noted that the background velocity and flow height in
this treatment are in general stationary functions of space.
Extracting the inviscid terms only from Eq.~(\ref{metric}) by 
setting $\nu=0$, it should be easy to ultimately render these 
terms into a compact formulation that looks like~\cite{blv05}
\begin{equation}
\label{compact}
\prt_\alpha \left( {\mathsf{f}}^{\alpha \beta} \prt_\beta
f^{\prm}\right) = 0 , 
\end{equation}
in which the Greek indices run from $0$ to $1$, with the
identification that $0$ stands for $t$ and $1$ stands for $r$.
On inspecting the terms in the left hand side of Eq.~(\ref{metric}),
the symmetric matrix, 
\begin{equation}
\label{symmat}
{\mathsf{f}}^{\alpha \beta } = v_0
\begin{bmatrix}
1 & v_0 \\
v_0 & v_0^2 - gh_0
\end{bmatrix} \\ , 
\end{equation}
can be obtained. It is well known that 
in terms of the metric ${\mathsf{g}}_{\alpha \beta}$, the d'Alembertian
for a scalar in curved space, is given by~\cite{blv05}
\begin{equation}
\label{alem}
\triangle \psi \equiv \frac{1}{\sqrt{-\mathsf{g}}}
\prt_\alpha \left({\sqrt{-\mathsf{g}}}\, {\mathsf{g}}^{\alpha \beta} 
\prt_\beta \psi \right) , 
\end{equation}
in which $\mathsf{g}^{\alpha \beta}$ is the inverse of the matrix
implied by ${\mathsf{g}}_{\alpha \beta}$.
Under the equivalence that ${\mathsf{f}}^{\alpha \beta } =
\sqrt{-\mathsf{g}}\, {\mathsf{g}}^{\alpha \beta}$, and therefore,
$\mathsf{g} = \det \left({\mathsf{f}}^{\alpha \beta }\right)$, an
effective metric can immediately be set down as
\begin{equation}
\label{effmet}
\mathsf{g}^{\alpha \beta}_{\mathrm{eff}} = 
\begin{bmatrix}
1 & v_0 \\
v_0 & v_0^2 - gh_0
\end{bmatrix} \\ , 
\end{equation}
which is entirely identical to the one derived by Sch\"{u}tzhold and
Unruh. By matrix inversion the inverse 
effective metric, $\mathsf{g}_{\alpha \beta}^{\mathrm{eff}}$,
is easily obtained from Eq.~(\ref{effmet}). For this system,
in which the spatial dependence is only on the radial coordinate, 
the critical condition, $v_0^2 = gh_0$, is also identified precisely as 
the horizon condition of either an acoustic black hole or an acoustic 
white hole, depending on the direction of the flow. 
The critical condition for a flow progressing from the super-critical 
region to the sub-critical region, is analogous to the horizon of an
acoustic white hole. 

In presenting the perturbative analysis in this paper, however, 
viscosity has been included in the governing equations, to finally
arrive at Eq.~(\ref{metric}). The presence of viscosity 
disrupts the precise symmetry of the inviscid conditions described by
Eq.~(\ref{compact}). This implies that the clear-cut horizon
condition, obtained from the inviscid limit, will be affected.
However, this effect, for small viscosity, as Sch\"{u}tzhold and Unruh
have argued, cannot be too drastic. They treated viscosity as a small 
adjunct effect on the inviscid flow, and concluded that the basic 
properties of gravity waves will not be affected overmuch. One way or 
the other, the most important feature to emerge from the analogy of a 
white hole horizon shall remain qualitatively unchanged, namely, that 
a disturbance propagating upstream from the sub-critical flow region 
(where $v_0^2 < gh_0$) cannot penetrate through the horizon into the 
super-critical region of the flow (where $v_0^2 > gh_0$). In keeping
with this, Volovik has also pointed out that the jump condition can 
be closely related to the horizon of a white hole~\cite{vol05}, a 
surface that nothing can penetrate. This property of the flow  
has a crucial bearing on the way the hydraulic jump is formed.

It has also been intriguing to note that the linearised perturbation 
dynamics in the shallow-water hydraulic jump is remarkably identical 
to the corresponding dynamics of astrophysical accretion on to an accretor 
(either a star or any compact object)~\cite{pso80,td92,ray03,crd06},
with the only important point of difference being that whereas in 
the former case a disturbance in the fluid propagates as surface  
gravity waves, in the latter case a disturbance propagates as acoustic 
waves. In all other important qualitative respects, the
similarities between what is an incompressible two-dimensional
hydraulic jump outflow taking place on the laboratory scale, and what
is a compressible spherically symmetric accretion inflow happening on
the astrophysical scale, are very much unexpected.
Apropos of this matter, 
however, Landau and Lifshitz have
made a general but very significant observation that ``there is a
remarkable analogy between gas flow and the flow in a gravitational
field of an incompressible fluid with a free surface, when the depth
of the fluid is small..."~\cite{landau}.

\section{Standing waves in the sub-critical region and scaling arguments}
\label{sec3}

The perturbation is treated as a standing wave, which is constrained 
to vanish at two separate boundaries. Between 
these two boundaries the flow has to be continuous. Since the jump 
is a discontinuity in the flow, the boundaries of the standing wave 
perturbation will have to be chosen on one side of the jump only, 
although Eq.~(\ref{metric}) itself holds true over the entire flow space. 
Therefore, the stability analysis is confined to the sub-critical region 
of the flow only, where $v_0 < c_{\mathrm g}$, with 
$c_{\mathrm g}$ being the steady value of the speed of surface
gravity waves, expressible as $c_{\mathrm g}^2 = gh_0$.
As for the boundaries of the 
perturbation, one of them can be the outer boundary of the steady 
flow itself, where, by virtue of the boundary condition on the steady 
flow, the perturbation would naturally decay out. The inner boundary 
on the other hand, may be chosen to be infinitesimally close to the 
jump, which can be conceived of as a standing shock front. Going by
the analogy of a white hole, this point will be seen as an impenetrable
surface by any disturbance in the sub-critical region, and, therefore, 
the amplitude of the standing wave will decay in its neighbourhood. 
Between these two boundaries (which 
actually define the entire sub-critical flow region) the flow would 
have entirely lost its laminar character, and it would be 
most suited to deriving some physical insight about the behaviour 
of the perturbation and the influence of viscosity on it. 

Upon using a solution of the form 
$f^{\prm}(r,t) = p_{\omega}(r) \exp(- {\mathrm i} \omega t)$, a quadratic 
equation in $\omega$ is obtained, and it is given by
\begin{equation}
- \omega^2 p_{\omega} - {\mathrm i} \omega 
\left [ 2 \frac{\mathrm d}{{\mathrm d}r}
\left (v_0 p_{\omega} \right ) - \nu \frac{p_{\omega}}{h_0^2} \right ] 
+ \frac{1}{v_0}  \frac{\mathrm d}{{\mathrm d}r} 
\left [v_0 \left ( v_0^2 - c_{\mathrm g}^2 \right )
\frac{{\mathrm d}p_{\omega}}{{\mathrm d}r} \right ] + 3 \nu 
\frac{v_0}{h_0^2} \frac{{\mathrm d}p_{\omega}}{{\mathrm d}r}=0 . 
\label{quadom}
\end{equation}
This expression is first multiplied by $v_0 p$ and, between the 
two chosen boundaries, an integration by parts is carried out. Imposing 
the requirement that all integrated ``surface" terms are to vanish at 
the two boundaries, will finally give 
\begin{equation}
\omega^2 \int v_0 p_{\omega}^2 \, {\mathrm d}r + {\mathrm i} 
\omega \nu \int \frac{v_0 p_{\omega}^2}{h_0^2} \, {\mathrm d}r
+ \int v_0 \left (v_0^2 - c_{\mathrm g}^2 \right ) 
\left (\frac{{\mathrm d}p_{\omega}}{{\mathrm d}r} \right )^2 \, 
{\mathrm d}r 
- 3 \nu \int \left (\frac{v_0}{h_0} \right )^2 p_{\omega} 
\frac{{\mathrm d}p_{\omega}}{{\mathrm d}r} \, {\mathrm d}r = 0 . 
\label{integom}
\end{equation}
A solution for $\omega$ can be found by solving the foregoing 
quadratic equation. Of immediate interest is the real part of the 
perturbation, going as 
\begin{equation}
\label{realpert}
\Re (- {\mathrm i}\omega) = - \frac{\nu}{2}  
\left[ \int \frac{v_0 p_{\omega}^2}{h_0^2} \, {\mathrm d}r \right] 
\left[\int v_0 p_{\omega}^2 \, {\mathrm d}r \right]^{-1} . 
\end{equation}
In the sub-critical region the flow height, $h_0 (r)$, is very weakly
dependent on $r$, and so $\Re(-{\mathrm i}\omega)\simeq -\nu /2h_0^2$. 
Under inviscid conditions, the perturbation will have a constant amplitude 
in time, but viscosity, with all its dissipative implications, causes the 
perturbation to decay out. This decay of the amplitude 
of the perturbation will go as $\exp(- \nu t/2 h_0^2)$, and it will 
also set up a time scale for viscosity to have its dissipative effect, 
which, to an order-of-magnitude, will be given 
by $t_{\mathrm{visc}} \sim h_0^2 / \nu$. 

The time scale on which viscous drag in the fluid will cause a 
dissipative slowing down of the flow, will also be given 
by $t_{\mathrm{visc}}$.
The information of an advanced layer of fluid slowing down has to
propagate upstream to preserve the smooth continuity of the fluid flow.
However, this propagation can happen no faster than the speed of
surface gravity waves, $c_\mathrm{g}$, and in the region
where $v_0 > c_\mathrm{g}$, no information, therefore, can propagate
upstream~\cite{landau,bpw97}. So a stream of fluid that has arrived
later, moves on ahead, unhindered and uninformed, till
its speed becomes comparable with the speed of the surface gravity waves,
and only then does any information about a ``barrier" ahead catches up
with the fluid. At this stage one may define a dynamic time scale,
$t_{\mathrm{dyn}} \sim {r}/{v_0}$, which is the
time scale on which the bulk flow proceeds.
Setting $t_{\mathrm{visc}} \simeq t_{\mathrm{dyn}}$ with the additional
requirement that $v_0 \simeq c_\mathrm{g}$, gives a condition for the
``news" of the viscous slowing down finally catching up with the bulk
flow itself. The continuity equation gives a further constraint that
$rv_0 h_0 = Q/2 \pi$, and on using this, a scaling behaviour for the 
hydraulic jump radius,
\begin{equation}
\label{jump}
r_{\mathrm{j}} \sim Q^{5/8} {\nu}^{-3/8} g^{-1/8} ,
\end{equation}
is obtained. This is exactly the scaling relation obtained by 
Bohr et al.~\cite{bdp93}, a result that was 
experimentally corroborated later by Hansen et al.~\cite{hansen97}.

The crucial factor that emerges from the above discussion is that for the
formation of the hydraulic jump, the two time scales, $t_{\mathrm{visc}}$
and $t_{\mathrm{dyn}}$, would have to match each other closely, when
the Froude number, $\mathcal{F}$, is close to unity.
Under these conditions, a layer of fluid arriving late is
confronted by a barrier formed by a layer of fluid moving ahead with
an abrupt slowness. This slowly moving layer of fluid flowed past 
earlier in time, and at far radial distances
it has been retarded considerably by viscous drag. In this
situation there could not be an indefinite accumulation of the fluid,
and since continuity of the fluid flow has to be preserved, the newly
arrived fluid layer slides over the earlier viscosity dragged slowly
moving layer of fluid, and what is seen is a sudden increase in the
height of the fluid layer, a phenomenon that is known as a hydraulic
jump. It is conceivable that this picture also explains the eddies which
form immediately after the jump~\cite{hansen97}.

\section{Standing waves in the super-critical region and the role of
nonlinearity} 
\label{sec4} 

So far the mathematical treatment was concerned primarily with 
developing and explaining a linearised perturbative effect on a 
steady background solution of the flow. However, what is actually 
being studied is a fluid dynamical system after all, and to derive
a complete understanding of this flow system, at one stage or the
other, it should be necessary to address the issue of nonlinearity.
It has been established through some earlier experimental and 
theoretical works~\cite{vol06,us05} that the stability of the 
flow in the super-critical region comes under threat, and in this
connection it is generally recognised that in one way or the other
nonlinearity has an involvement. 

In this context
the question that has to be first posed is about the possibility 
of carrying out a self-consistent perturbative analysis that 
encompasses nonlinearity. Going back to the earlier prescription
of $v(r,t) = v_0(r) + v^{\prm}(r,t)$ and 
$h(r,t) = h_0(r) + h^{\prm}(r,t)$, but retaining the nonlinear 
term in the definition of $f^{\prime}$, it is easy to show that 
$f^{\prm}=r\left(v_0 h^{\prm}+h_0v^{\prm}+v^{\prm}h^{\prm}\right)$,
while 
\begin{equation}
\label{nlfluch}
\frac{\prt h^{\prm}}{\prt t} = - \frac{1}{r}
\frac{\prt f^{\prm}}{\prt r}
\end{equation}
and 
\begin{equation}
\label{nlflucv}
\frac{\prt v^{\prm}}{\prt t} = \frac{1}{rh}\left(\frac{\prt f^{\prm}}
{\prt t}\right) + \frac{v}{rh}
\left(\frac{\prt f^{\prm}}{\prt r} \right) .
\end{equation}
The two foregoing results are mathematically exact, and on comparing 
them with Eqs.~(\ref{fluch}) and~(\ref{flucv}), they are seen to be
rather intriguing
too. They show that regardless of the nature of the background flow 
--- stationary or otherwise --- the mathematical interdependence among
$v^{\prm}$, $\rho^{\prm}$ and $f^{\prm}$ retain an invariant form. 
This leads, as it shall be presently demonstrated, to some interesting
consequences so far as nonlinearity is concerned. Following the 
principle behind the derivation of Eq.~(\ref{metric}), it can be 
shown that on retaining the lowest order of nonlinearity it is 
possible to arrive at a similar relation that is of the form
\begin{equation}
\left({\mathcal L}_0 + \lambda {\mathcal L}_1 \right)f^{\prm} = 
- \nu \frac{v_0}{h_0^2} \left\{ \left(1 - 3 \lambda 
\frac{h^{\prm}}{h_0} \right) \frac{\prt f^{\prm}}{\prt t} + 3v_0
\left[ 1 + \lambda \left(\frac{v^{\prm}}{v_0} - 3 \frac{h^{\prm}}{h_0}
\right) \right] \frac{\prt f^{\prm}}{\prt r} \right\} ,
\label{nlmetric} 
\end{equation}
in which ${\mathcal L}_0$ is the operator implied 
by Eqs.~(\ref{compact}) and~(\ref{symmat}), and $\lambda$ is a 
``switch" parameter which can assume values of $0$ and $1$ only, 
depending, respectively, on whether one does not or does want to 
retain nonlinearity in the lowest order. The most interesting aspect
of Eq.~(\ref{nlmetric}), however, is to be seen in ${\mathcal L}_1$, 
which is an operator that derives solely from accounting for lowest-order
nonlinearity (i.e. with fluctuations upto the second order only). The 
structure of ${\mathcal L}_1$ is exactly similar to the form of 
${\mathcal L}_0$, as Eq.~(\ref{compact}) gives it, with the only 
difference being that the corresponding metric elements, referred 
to in this case as ${\tilde{\mathsf{f}}}^{\alpha \beta}$, will involve
time-dependent first-order fluctuations, unlike the case of the 
stationary background state, given by Eq.~(\ref{symmat}). Expressed
fully these metric elements will read as 
\begin{equation}
\label{nlsymmat}
{\tilde{\mathsf{f}}}^{\alpha \beta} = -v_0
\begin{bmatrix}
{\tilde h} & v_0 {\tilde h} - v^{\prm}  \\
v_0 {\tilde h} - v^{\prm} & 
v_0^2 {\tilde h} - 2v_0 v^{\prm}
\end{bmatrix} \\ ,
\end{equation}
in which ${\tilde h} = h^{\prm}/h_0$, and the Greek indices,
$\alpha$ and $\beta$, both run from $0$ to $1$, as before. 
The fundamentally symmetric 
form of ${\tilde{\mathsf{f}}}^{\alpha \beta}$ can only mean that 
even with nonlinearity incorporated in the treatment, in the inviscid
limit (i.e. for $\nu = 0$), it should be very much possible to construct
an analogue acoustic black hole metric (or white hole metric, depending 
on the direction of the flow), following the mathematical procedure 
adopted earlier in Section~\ref{sec2}. Going by the form of 
Eqs.~(\ref{nlfluch}) and~(\ref{nlflucv}), it is also worth surmising
at this point that this invariance will not be disturbed even when 
nonlinearity  upto any higher order might be accounted for. 

Another aspect of Eq.~(\ref{nlmetric}) is that apart from nonlinearity
it also incorporates viscosity. The question is to what extent viscosity
is important in determining the fate of a standing wave perturbation in
the super-critical region. Going back to Eq.~(\ref{integom}), but in 
this instance imposing the requirement that the standing wave will be
bounded within the super-critical region only, close to the jump itself
and at a point very close to the origin of the radial flow, it will be 
easy to see that for $v_0 > c_{\mathrm{g}}$, there will be a growing
mode for the perturbation. A condition for this instability can be 
set down as 
\begin{equation}
\label{instabcon}
\int \bigg[ v_0 \left(v_0^2 - c_{\mathrm g}^2 \right)
\left(\frac{{\mathrm d}p_{\omega}}{{\mathrm d}r}\right)^2
- 3\nu \left(\frac{v_0}{h_0}\right)^2 p_{\omega}
\frac{{\mathrm d}p_{\omega}}{{\mathrm d}r}\bigg] \,{\mathrm d}r >0 , 
\end{equation}
whose form implies that if viscosity, quantified by the value of $\nu$,
is large enough, then the instability could be overcome. To this 
extent viscosity might be seen to oppose instability in the 
super-critical region. However, this conclusion is derived solely 
within the linearised framework, and what is more, in the 
super-critical region the flow is largely laminar till the jump 
position is reached. Therefore, the true physical role of viscosity 
may not be of much significance in this region, which is dominated
much more by nonlinearity. In that event, so far as the standing 
wave in the super-critical region is concerned, it would be reasonable
to drop viscosity (by setting $\nu= 0$) in Eq.~(\ref{nlmetric}), and
retain only the time-dependent nonlinear effects on the background 
linearised stationary state. 

Having done so, employing the same mathematical methods and physical
arguments that led to Eq.~(\ref{integom}), 
and making use of the fact that 
${\tilde{\mathsf{f}}}^{01}= {\tilde{\mathsf{f}}}^{10}$, a quadratic 
dispersion relation in $\omega$ can be derived from 
Eqs.~(\ref{nlmetric}) and~(\ref{nlsymmat}), which will read as 
\begin{equation}
\label{nlintegom}
\omega^2 \int p_{\omega}^2 \left(v_0 + \lambda 
{\tilde{\mathsf{f}}}^{00}\right) \, {\mathrm d}r + {\mathrm i}\lambda
\omega \int p_{\omega}^2 \frac{\prt {\tilde{\mathsf{f}}}^{00}}{\prt t}\,
{\mathrm d}r + \int v_0 \left (v_0^2 - c_{\mathrm g}^2 \right )
\left(\frac{{\mathrm d}p_{\omega}}{{\mathrm d}r}\right)^2 \,
{\mathrm d}r - \lambda \int \bigg[ p_{\omega}
\frac{{\mathrm d}p_{\omega}}{{\mathrm d}r} 
\frac{\prt {\tilde{\mathsf{f}}}^{01}}{\prt t} - 
{\tilde{\mathsf{f}}}^{11} 
\left(\frac{{\mathrm d}p_{\omega}}{{\mathrm d}r}\right)^2\bigg]\,
{{\mathrm d}r}=0 .  
\end{equation}
It will be easy to conclude from this expression that the time-dependent 
part of the perturbation, given by $\exp(- {\mathrm i} \omega t)$, will 
have no complex part and, therefore, there will be a divergent mode of 
the perturbation arising solely due to nonlinearity. 

While all this can be deduced about the amplitude-dependent phase (a 
regular feature where nonlinearity is concerned), it does not indicate 
anything precise about the behaviour of the amplitude of the perturbation 
itself. To do so it shall be necessary
to adopt an approach that is quite common in studying amplitude 
equations~\cite{stro,js99}. First the expression for $f^{\prm}(r,t)$
is to be modified slightly and written as 
$f^{\prm}(r,t) = \xi(t) p_{\omega}(r) \exp(- {\mathrm i} \omega t)$,
with $\xi(t)$ being a slowly-varying function of time, containing
information on the nonlinear time-dependent growth of the amplitude. 
The behaviour of $\xi(t)$ is to be studied at a fixed value of $r$
very close to the jump itself, where, as experimental evidence shows,
the amplitude is saturated by nonlinear effects~\cite{vol06}. The 
justification for imposing the requirement of $\xi(t)$ being a 
slowly-varying function of time lies in the fact that the time scale
for the amplitude amplification of the perturbation in the vicinity of 
the jump is much greater than the dynamic time scale for the formation of
the jump, $r_{\mathrm{j}}/c_{\mathrm{g}}$. 

Now Eqs.~(\ref{nlfluch}) and~(\ref{nlflucv}) connect $h^{\prm}$ and
$v^{\prm}$, respectively, to $f^{\prm}$. Considering the modified
expression of $f^{\prm}$, it should therefore be easy to appreciate
that both $h^{\prm} \sim \xi(t)$ and $v^{\prm} \sim \xi(t)$. By this 
argument, since $\omega$ also contains $h^{\prm}$ and $v^{\prm}$,
it should be noted that $\omega \sim \xi(t)$. Following this 
understanding, using the modified form of $f^{\prm}(r,t)$ in 
Eq.~(\ref{nlmetric}), and neglecting all second derivatives of 
$\xi(t)$, it can now be argued that the general form of the variation 
of $\xi(t)$ with respect to $t$ will be given by 
${\dot{\xi}}=a \xi + b \xi^3$, with any even power of $\xi(t)$
being left out to preserve time-reversal symmetry, when 
$\xi \longrightarrow - \xi$. The coefficients $a$ and $b$ 
will determine the growth of $\xi$, and with $a>0$ there will be an
early growth mode, but this growth will be saturated if $b<0$. That 
this is precisely what happens inside the super-critical region 
close to the jump, is very much in evidence through experimental 
studies~\cite{vol06}. With regard to this it would also be quite worthwhile
to draw attention to the fact that for a shallow-layer flow of liquid
helium, below its superfluid transition temperature, a very striking
feature is the occurrence of a standing capillary wave (stationary
ripples) between the impact point of the jet and the jump 
radius~\cite{rol07}. 

\section{Travelling waves} 
\label{sec5}

In treating the perturbation as radially propagating waves of very high 
frequency, a condition that is needed to be satisfied is that 
the frequency, $\omega$, of the travelling waves should be much greater
than any characteristic frequency in the system. Since $t_{\mathrm{visc}}$ 
appears as a natural time scale in the system, it would be logical to use 
this time scale and impose the condition that 
$\omega \gg t_{\mathrm{visc}}^{-1}$. The spatial part of the 
perturbation is prescribed as $p_\omega (r)=e^s$, with 
$s$ itself being given by the power series, 
\begin{equation}
\label{powser}
s(r) = \sum_{n=-1}^{\infty} \omega^{-n} k_n(r) . 
\end{equation}
The complete perturbation will
be given as $f^{\prm}(r,t)=\exp(s-{\mathrm i}\omega t)$. Following 
some algebra, Eq.~(\ref{quadom}) can be written in a slightly altered 
form as 
\begin{equation}
\left (v_0^2 - c_{\mathrm g}^2 \right ) \frac{{\mathrm d}^2 p_\omega}
{{\mathrm d}r^2} 
+ \left [ 3 v_0 \frac{{\mathrm d} v_0}{{\mathrm d}r}
- \frac{1}{v_0}\frac{{\mathrm d}}{{\mathrm d}r} \left (v_0 
c_{\mathrm g}^2 \right ) - 2 {\mathrm i} v_0 \omega 
+ 3 \nu \frac{v_0}{h_0^2} \right ] 
\frac{{\mathrm d}p_\omega}{{\mathrm d}r} 
- \omega \left ( \omega + 2 {\mathrm i} 
\frac{{\mathrm d} v_0}{{\mathrm d}r} + {\mathrm i} \nu 
\frac{1}{h_0^2} \right ) p_\omega = 0 , 
\label{recast}
\end{equation}
in which $p_\omega (r)$ is substituted by making use of the power series 
given by Eq.~(\ref{powser}). After this the three successive 
highest-order terms in $\omega$ will be obtained as $\omega^2$, $\omega$
and $\omega^0$, all of whose coefficients are collected and summed up 
separately, and each individual sum of coefficients is then set to zero. 
Solutions resulting from expressions involving $\omega^2$ and $\omega$, 
are given as
\begin{equation}
\label{kayminus1}
k_{-1} = \int \frac{\mathrm i}{v_0 \pm c_{\mathrm g}} \, {\mathrm d}r
\end{equation}
and
\begin{equation}
\label{kaynot}
k_0 = -\frac{1}{2}\ln\left(v_0 c_{\mathrm g} \right)
\pm \nu \int \frac{1}{c_{\mathrm g} h_0^2} \left [ 1 \mp 
\frac{3}{2 \left ( {\mathcal F} \pm 1 \right )} \right ] \, 
{\mathrm d}r , 
\end{equation}
respectively, with ${\mathcal F}$ being the Froude number, 
${\mathcal F}=v_0/c_{\mathrm g}$. 

It will be necessary to show that all successive terms 
of $s(r)$ will self-consistently follow the condition
$\omega^{-n}\vert k_n(r)\vert \gg \omega^{-(n+1)}\vert k_{n+1}(r)\vert$,
i.e. the power series given by $p_\omega (r)$ will converge very rapidly
with increasing $n$, so that the series can be truncated after 
the first few terms. This whole approach of studying high-frequency 
travelling waves, has also been successfully applied in astrophysical 
flows~\cite{pso80,crd06}. The convergence criterion can be shown
to be very much true for the radial shallow-water flow here, 
considering the behaviour of the flow 
in both the super-critical (where $v_0 \sim \mathrm{constant}$ and 
$h_0 \sim r^{-1}$) and the sub-critical (where $v_0 \sim r^{-1}$ and 
$h_0 \sim \mathrm{constant}$) regions. In the inviscid limit the first 
three terms in $k_n(r)$ will behave  
as $k_{-1} \sim r$, $k_0 \sim \ln r$ and $k_1 \sim r^{-1}$. With the
inclusion of viscosity as a physical effect, it can be seen from
Eqs.~(\ref{kayminus1}) and~(\ref{kaynot}), respectively, that while
$k_{-1}$ remains unaffected, $k_0$ acquires a $\nu$-dependent term that 
goes asymptotically as $r$. Similarly, two such viscosity-dependent 
terms, going as $r$ and $\ln r$, respectively, are added to $k_1$.
However, since $\nu$ has a relatively small value for water, and since 
the frequency of the travelling waves is at the same time 
very great (implied by $\omega \gg t_{\mathrm{visc}}^{-1}$), the 
self-consistency requirement still holds. Therefore, it should be quite 
sufficient to truncate the power series expansion of $s(r)$ after 
considering the two leading terms only, and leave out $k_1$, which will 
anyway contibute only to the phase of the perturbation and not its 
amplitude. An expression for the perturbation will, therefore, be 
obtained as
\begin{equation}
f^{\prm}(r,t) \simeq \frac{A_\pm}{\sqrt{v_0 c_{\mathrm g}}}
\exp \left [ \int \left ( \frac{{\mathrm i} \omega}{v_0 \pm c_{\mathrm g}}
\pm \frac{\nu}{c_{\mathrm g} h_0^2} \left [ 1 \mp
\frac{3}{2 \left ( {\mathcal F} \pm 1 \right )} \right ] 
\right ) \, {\mathrm d}r  \right ] e^{-{\mathrm i} \omega t} , 
\label{fpertur}
\end{equation}
which should be seen as a linear superposition
of two waves with arbitrary constants $A_+$ and $A_-$. Both these
two waves move with a velocity, $c_{\mathrm g}$, relative to the fluid,
one against the bulk flow and the other along with it, while the bulk 
flow itself has a velocity, $v_0$.  
All questions pertaining to the growth or decay in the amplitude of 
the perturbation will be crucially decided by the real terms derived 
from $k_0$. For the choice of the upper sign, i.e. for outgoing waves, 
the amplitude will be amplified as the wave propagates for a certain 
distance outwards, and then it will start decaying. A quantitative 
estimate can be made of the asymptotic behaviour of the perturbation. 
For the highly super-critical region of the flow, much before the jump, 
$v_0$ is approximately constant, while $h_0 \sim r^{-1}$. 
The inviscid term in Eq.~(\ref{kaynot}) will thus contribute a growth 
behaviour going as $r^{1/4}$. The growth of the viscous term on the 
other hand will vary as  
$\exp \left[(r/r_{\mathrm{s1}})^{3.5} \right]$, with 
$r_\mathrm{s1}$ being a suitable scale factor. At the opposite end,
when the flow is in the highly sub-critical region, $h_0$ will 
be nearly constant, while $v_0 \sim r^{-1}$. In this
region the contribution of the inviscid term to the perturbation 
will be in the form of a growth going as $r^{1/2}$, which, however, 
will be suppressed by an exponential decay contributed by the viscous
term, going approximately as $\exp (-r/r_{\mathrm{s2}})$, with 
$r_{\mathrm{s2}}$ being another scale factor. Interestingly 
enough, in the vicinity of the jump, the scale factor, $r_{\mathrm{s2}}$, 
reproduces the exact scaling form of the jump radius derived 
by Bohr et al.~\cite{bdp93}. Quite evidently, the 
perturbation starts decaying after crossing the region where 
$\mathcal{F} = 0.5$. However, even when the outgoing wave grows in
amplitude, the growth is not unrestrained anywhere in the region of 
physical interest, and although it might disturb the jump itself 
somewhat while passing through it, the stability of the flow as 
well as the jump will not be too much adversely affected overall. 
So all that happens in the case of the outgoing wave is that 
starting from the super-critical region, it is completely 
transmitted through the jump discontinuity, continuing to grow 
in amplitude for some distance even beyond the jump, but decaying 
out eventually at very large distances. 

This relatively placid state of affairs receives a severe jolt 
with the choice of the lower sign in Eq.~(\ref{kaynot}), i.e. for 
incoming waves. It is evident from Eq.~(\ref{kaynot}) that for
an incoming wave progressing upstream from the highly sub-critical
flow region to the vicinity of the jump (where the Froude number,
$\mathcal F$, will be close to unity), the viscous term will cause 
a large exponential divergence in the amplitude of the perturbation, 
and the flow will therefore suffer noticeably from large 
fluctuations. On the other hand, in the super-critical region, 
it is because of the viscous term once again that the flow will be 
pronouncedly stable under any perturbation propagating upstream.
Physically speaking this is what it should be anyway, because in 
the super-critical region the bulk flow proceeds much faster 
against the propagation of the perturbation,
and since the super-critical region itself has only a finite
spatial extent, the perturbation is not allowed to remain here
for a span of time long enough to destabilise the flow in any
way. Indeed, in the super-critical region just before the jump, the 
amplitude of the perturbation decays to vanishingly small values.
This can only imply that any incoming perturbation propagating 
upstream from the sub-critical region is not allowed to be 
transmitted through the jump discontinuity into the super-critical
region, and this is entirely in agreement with the white hole analogy
presented at the end of Section~\ref{sec2}. 

\begin{figure}[t]
\begin{center}
\includegraphics[scale=1.0, angle=0]{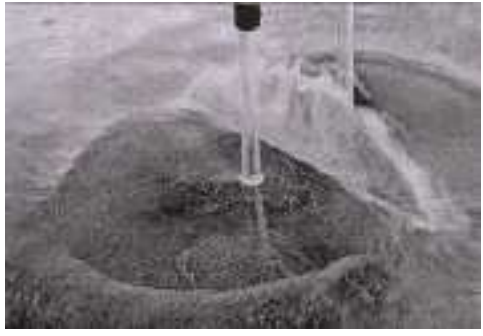}
\caption{\label{f1} \small{An oblique view of the wall of water
raised between two adjacent hydraulic jumps (Photo obtained by
the courtesy of R. P. Kate, P. K. Das and S. Chakrabarti).}}
\end{center}
\end{figure}

This theoretical contention has received much support from an 
experimental work carried out by Kate et al.~\cite{kdc06} to study 
hydraulic jumps formed due to adjacent normal impinging liquid jets. 
In a series of clearly
illustrative photographs, these authors have recorded that when one
of the two adjacent liquid jets is gradually moved along a straight
line towards the other one, there will come a stage when, with the 
circular jumps formed by the two vertical jets being close enough to
each other, the
liquid trapped between the two jumps will be noticeably pushed up to
stand at a much greater height than the rest of the flow. In a manner
of speaking, a small wall of water --- the liquid used in this 
experimental study --- will be formed. It is easy to see that the 
jump formed by the moving jet behaves as a sub-critical disturbance
propagating upstream towards the jump formed due to the static jet. 
This disturbance cannot penetrate through the static jump, and so 
as it approaches the jump region, its vertical height increases 
dramatically, due to accumulation of the liquid. This point has been
amply demonstrated by the photograph (taken by Kate et al.~\cite{kdc06})
in Figure~\ref{f1}. However, the disturbance propagating upstream in
this experiment, is not a purely 
axisymmetric disturbance, and hence, there is a lack of symmetry in 
the formation of the wall of water between the two jumps. 
It is conceivable that if 
the perturbation propagating upstream were constrained to have been 
perfectly axisymmetric in nature, then one would have obtained a 
manifestly circular wall of water to be caught between the incoming
waves and the periphery of the circular hydraulic jump. Relaxation 
for this situation can be achieved through the perturbation expending 
itself in destabilising the jump. That the flow rate (and connected 
to it by a scaling relation, 
the jump position itself) is highly destabilised by a travelling 
perturbation incident from the sub-critical side, is a feature
that is in quite close correspondence with the observation 
of Hansen et al.~\cite{hansen97} that the jump position starts 
fluctuating beyond a certain critical value of the flow rate.   

The time-averaged energy flux associated with the travelling 
perturbation can be estimated by first noting that in an annular
element of fluid, the energy arising due to the perturbation 
is given by the sum of its kinetic energy, 
\begin{equation}
\label{ekin}
{\mathcal E}_{\mathrm{kin}} = 2 \pi r \frac{\rho}{2}
\left( h_0 + h^{\prm} \right) \left (v_0 + v^{\prm} \right )^2 \,
{\mathrm d}r , 
\end{equation}
and its internal energy,
\begin{equation}
\label{eint}
{\mathcal E}_{\mathrm{int}} = 2 \pi r \rho \left [ \epsilon h_0
+ h^{\prm} \frac{\prt}{\prt h_0}\left(h_0 \epsilon \right) +
\frac{1}{2}{h^{\prm}}^2 \frac{\prt^2}{\prt h_0^2}
\left (h_0 \epsilon \right) \right ] \, {\mathrm d}r , 
\end{equation}
in which $\rho$ is the constant density of the fluid, and $\epsilon$ 
is the internal energy per unit mass~\cite{landau}.
The first-order terms all vanish upon time-averaging, leaving an
expression for the energy of the perturbation that involves only
second-order terms. This reads as
\begin{equation}
\label{epert}
{\mathcal E}_{\mathrm{pert}} = 2 \pi r \frac{\rho}{2}
\left [ h_0 {v^{\prm}}^2 + 2 v_0 h^{\prm} v^{\prm} +
{h^{\prm}}^2 \frac{\prt^2}{\prt h_0^2}
\left (h_0 \epsilon \right) \right ] \, {\mathrm d}r . 
\end{equation}
The height fluctuation, $h^{\prm}$, and the velocity fluctuation, 
$v^{\prm}$, could both be expressed in terms of $f^{\prm}$, by making
use of Eqs.~(\ref{fluch}) and~(\ref{flucf}), respectively. Carrying out 
a time-averaging over the periodic part of the square of $f^{\prm}$, 
under the approximation that to a leading order $s \simeq \omega k_{-1}$, 
will lead to an expression for the energy flux of the propagating 
perturbation as
\begin{equation}
{\mathrm F}_{\mathrm{pert}} \simeq \frac{\pi^2 \rho}{Q}
\frac{A_\pm^2}{{\mathcal F} \pm 1} \left [1 \pm 2 {\mathcal F} +
\frac{1}{g}\frac{\prt^2}{\prt h_0^2} \left (h_0 \epsilon \right)
\right ] 
\exp \left ( \pm 2 \nu \int \frac{1}{c_{\mathrm g} h_0^2}
\left [1 \mp \frac{3}{2 \left ( {\mathcal F} \pm 1 \right )} \right ]
\, {\mathrm d}r \right ) , 
\end{equation}
\label{flux}
with a factor of $1/2$ deriving from the time-averaging. 
Once again it can be seen that the incoming wave will cause
much destabilisation for the energy flux in the region where the 
value of the Froude number, $\mathcal{F}$, approaches unity. 

\section{Concluding remarks}
\label{sec6}

Some general comments would well be in order here, considering the 
broad appeal that the physical problem of the hydraulic jump holds
across various apparently unrelated disciplines. In this paper a
physical explanation for the hydraulic jump has been furnished with
the help of the analogy of an acoustic white hole, and by applying
standard mathematical tools from stability studies in astrophysical
accretion. In doing so the most salient fact that has been unearthed
is that the invariance required to invoke the analogy of acoustic 
geometry holds good even when nonlinearity is taken into account. 
So this whole study has amounted to a very striking convergence of 
fluid dynamics (both laboratory and astrophysical), analogue gravity
and the mathematics of nonlinear phenomena. 

The reverse aspect of this condition is also true. 
Recent literature has found mention of astrophysical problems
being addressed in terms of the analogy of a hydraulic jump. The
radial profiles of large gaseous baryonic structures, such as
galaxies and galaxy clusters, follow one power law in the inner
part, and a different one in the outer part, beyond a certain
characteristic radius. This transition has been seen to be sharp.
It does not necessarily imply that the two distinctly different
profiles are premised on different kinds of physical properties.
Rather, very much like the hydraulic jump, the governing equations
will have two solutions, and both will be availed of
simultaneously~\cite{hans03}. Another related astrophysical 
connection pertains to spiral shocks in protoplanetary discs. 
Spiral shocks, for most protoplanetary discs, create 
a loss of vertical equilibrium in the post-shock region, and 
this results in a rapid expansion of the gas in a direction that 
is perpendicular to the plane of the disc. This expansion has 
characteristics that are similar to hydraulic jumps~\cite{bol06}. 

While these facts are by now well established, the present work 
may also broach a related issue. In astrophysics, the question
of merging black holes is presently being much pondered over. 
Some light could be cast on this question by the kind of 
experiment conducted by Kate et al.~\cite{kdc06}, which has 
provided support for some of the theoretical arguments in this 
paper. The collision of two adjacent circular hydraulic jumps, 
both modelled as acoustic white holes, could provide a laboratory 
test case for the cognate situation of merging astrophysical 
black holes.  

\begin{acknowledgments}
The authors are grateful to Prasanta K. Das, Rama Govindarajan 
and Steen H. Hansen for many helpful comments. The authors are 
also indebted to Tapas K. Das for drawing attention to some 
recent works in the subject of analogue gravity, and to Ajit 
K. Kembhavi for extending unstinted support in various respects.
\end{acknowledgments}

\bibliography{pla_rev2}

\end{document}